\documentclass[10pt,headings=small,reqno,twocolumn]{scrartcl}
\usepackage{amsmath}
\usepackage{amsfonts}
\usepackage{amssymb}
\usepackage{latexsym}
\usepackage{graphicx}
\usepackage{epsf}

\allowdisplaybreaks \numberwithin{equation}{section}

\newcommand{\be}{\begin{equation}}
\newcommand{\ee}{\end{equation}}
\newcommand{\bi}{\begin{itemize}}
\newcommand{\ei}{\end{itemize}}
\newcommand{\bea}{\begin{eqnarray}}
\newcommand{\eea}{\end{eqnarray}}

 \let\b=\beta    
        
\let\m=\mu    \let\n=\nu

\let\==\equiv

\newcommand{\cD}{\mathcal{D}}

\newcommand{\cR}{\mathcal{R}}

\newcommand{\lb}{\bar{\lambda}}

\newcommand{\e}{\mathrm{e}}
\newcommand{\p}{\partial}

\begin{document}

\title{\Large{The $R^2$ phase-diagram of QEG 
and its spectral dimension}}

\author{\normalsize{Stefan Rechenberger and Frank Saueressig}\\ \small\it{Institute of Physics, University of Mainz, Staudingerweg 7, D-55099 Mainz, Germany.}\\ \small E-mail: \it{rechenbe@thep.physik.uni-mainz.de, saueressig@thep.physik.uni-mainz.de}}

\date{}

\twocolumn[
\begin{@twocolumnfalse}
\maketitle
\begin{abstract}
\noindent
Within the gravitational asymptotic safety program, the RG flow of the $R^2$ truncation in three and four spacetime dimensions is analyzed in detail. In particular, we construct RG trajectories which emanate from the non-Gaussian UV fixed point and possess long classical regimes where the effective average action is well approximated by the classical Einstein-Hilbert action. As an application we study the spectral dimension of the effective QEG spacetimes resulting from these trajectories, establishing that the picture of a multi-fractal spacetime is robust under the extension of the truncated theory space. We demonstrate that regimes of constant spectral dimensions can either be attributed to universal features of RG fixed points or singular loci in the $\beta$ functions. 
\end{abstract}
\vspace{1cm}
\end{@twocolumnfalse}
]
\section{Introduction}
\label{Intro}
Weinbergs asymptotic safety scenario \cite{Weinberg:1980gg,Weinproc1} strives toward constructing a quantum theory of gravity based on a non-Gaussian fixed point (NGFP) of the gravitational renormalization group (RG) flow  (see \cite{Niedermaier:2006wt,robrev,Reuter:2012id} for reviews). This NGFP is supposed to control the behavior of the theory at high energies and renders it safe from unphysical divergences. In order to ensure the predictivity of the construction, the NGFP has to come with a finite number of UV-relevant directions, i.e., the space of RG trajectories attracted to the fixed point at high energies is finite-dimensional. Moreover one has the phenomenological requirement that at least one of the RG trajectories emanating from the NGFP possesses a ``classical limit'' where the Einstein-Hilbert action  constitutes a good approximation of the gravitational interaction. The resulting quantum theory for gravity is called Quantum Einstein Gravity (QEG).

A central element for investigating Asymptotic Safety is the functional renormalization group equation (FRGE) for the gravitational average action $\Gamma_k$ \cite{Reuter:1996cp}, which allows the systematic construction of non-perturbative approximations of the theories $\b$ functions. Projecting the RG flow to subspaces of successively increasing complexity, the existence of a NGFP has been established in the Einstein-Hilbert truncation \cite{souma,oliver1,frank1}, the $R^2$ truncation \cite{oliver3,Lauscher:2002sq}, $f(R)$ truncations \cite{Codello:2007bd}, and truncations including a Weyl-squared term \cite{Benedetti:2009rx}. Moreover, refs.\ \cite{ frank1,Fischer:2006at} studied the properties of the NGFP in spacetime dimension more than four, while the quantum effects in the ghost sector have been investigated in \cite{Eichhorn:2009ah}. Furthermore boundary terms are considered in \cite{Becker:2012js}. QEG with Lorentzian signature was investigated in \cite{Manrique:2011jc} and an computer based algorithm for evaluating the flow equations was proposed in \cite{Benedetti:2010nr}. Besides providing striking evidence for the existence of the NGFP underlying Asymptotic Safety, the works including higher derivative interactions also provided strong hints that the fixed point possesses a finite number of relevant directions.

\begin{figure}[t!]
\begin{center}
\includegraphics[width=0.4\textwidth]{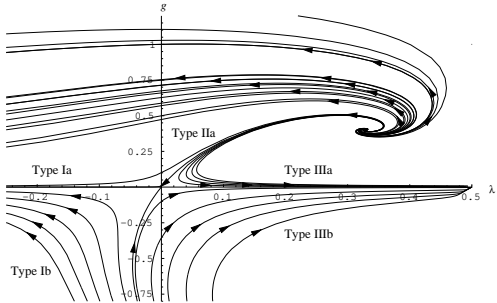}
\caption{Classification of the RG flow in the Einstein-Hilbert truncation \cite{Reuter:2001ag}.}
\label{fig:eh}
\end{center}
\end{figure}

While there is solid evidence in favor of a NGFP in quantum gravity, a lot less is known about the IR physics of the RG trajectories emanating from it.
In case of the Einstein-Hilbert truncation the RG trajectories have been classified in \cite{frank1} (also see \cite{Litim:2012vz} for a recent account) and are shown in Fig.\ \ref{fig:eh}.
All trajectories with positive Newtons constant are attracted to the NGFP in the UV and develop a classical regime close to the $\lambda$ axis. 
Based on this classification, the RG trajectory realized by Nature has been identified in \cite{h3}.

Since the $\b$ functions of more sophisticated truncations quickly become very involved, there is only very little known about its dynamics of the RG flow beyond the
Einstein-Hilbert case, however. One aim of the present work is to fill this gap by constructing families of trajectories originating from the full flow equations of the $R^2$ truncation in $d=3$ and $d=4$ spacetime dimensions.

Our motivation for this study is two-fold. With respect to the fundamental aspects underlying Asymptotic Safety, it is important to understand to which extend the features of the RG flow shown in Fig.\ \ref{fig:eh} remain robust when the effect of additional running couplings are included. In particular the existence of a classical regime can only be established by studying the dynamics of the RG flow and the interplay of different fixed points. Surprisingly, our analysis of the $R^2$ truncation also provides further insights on the IR fixed point recently discussed in \cite{Donkin:2012ud,Nagy:2012rn,Litim:2012vz}.

A more phenomenological motivation arises from studying quantum effects in gravitational systems via an ``RG improvement'', taking into account the scale-dependence of the gravitational coupling constants when studying, e.g., black holes \cite{reu:bh,reu:erick1}, cosmology \cite{reu:cosmo1,reu:entropy,reu:cosmo2,reu:elo,reu:wein3,reu:h1,reu:Ward:2008sm,reu:Bonanno:2010bt,Hindmarsh:2011hx}, or galaxy rotation curves \cite{reu:h2}. The RG improvement requires rather explicit knowledge of the RG trajectories underlying the physical system. Thus our results may be used to investigate the robustness of previous studies.

As a first application in this direction, we extend the formalism \cite{Lauscher:2005qz,Reuter:2011ah} to study the multi-fractal properties of the effective QEG spacetimes captured by the spectral dimension $D_s$. The interesting question from the functional renormalization group point of view is whether the fractal properties found within the Einstein-Hilbert truncation \cite{Reuter:2011ah} stay valid in higher truncations and if there are new features special to the higher derivative ansatz. In this sense, the $R^2$ truncation constitutes a natural choice for computing on shell corrections to the fractal properties of spacetime observed in the Einstein-Hilbert case. 

Notably, the interest in spectral dimension of spacetime is not limited to the functional renormalization group, since it is also accessible within other approaches to quantum gravity, as e.g., causal dynamical triangulation (CDT) \cite{Ambjorn:2005db}, euclidean dynamical triangulation (EDT) \cite{Coumbe:2012qr}, loop quantum gravity and spin foam models \cite{Modesto:2008jz} or in the presence of a minimal length scale \cite{Modesto:2009qc}. Studying the spectral dimension from various different angles may unravel universal features in the quantum structure of spacetime. On more phenomenological grounds one can build toy models of spacetime which encompass special features of the spectral dimension \cite{Giasemidis:2012qk}. Among further developments a fractional differential calculus \cite{Calcagni:2011kn} was used in \cite{Calcagni:2009kc,Arzano:2011yt} to assemble spacetimes with the fractal features observed in quantum gravity. Thus $D_s$ constitutes a  quite useful probe for quantum gravity effects.

The remaining parts of the paper are organized as follows. The phase diagrams of the $R^2$ truncation are constructed in section \ref{NGFPdata}. Section \ref{sec:diffusion} introduces the formalism for investigating the spectral dimension $D_s$ and we give explicit examples in section \ref{sec:spectral}. We close with a discussion of our results in section \ref{sec:conclusion}.

\section{Fixed points, phase diagrams and Asymptotic Safety}
\label{NGFPdata}
The gravitational part of the $R^2$ truncation takes the form
\be\label{ansatz}
\Gamma_k^{\rm grav}\!\!\! = \!\!\! \int \!\! d^dx \sqrt{g} \left[ \tfrac{1}{16 \pi G_k} \left( - R + 2 \bar{\lambda}_k \right) + \tfrac{1}{\bar{b}_k} R^2  \right] \, . 
\ee
The $\b$ functions governing the dependence of the dimensionful Newtons constant $G_k$, cosmological constant $\bar{\lambda}_k$ and $R^2$ coupling $\bar{b}_k$ on the RG scale $k$
 have been derived in \cite{Lauscher:2002sq} for general dimension $d$ and regulator $\cR_k$. In order to facilitate our numerical analysis, we will work with the optimized cutoff \cite{Litim:opt} throughout.
\subsection{$\b$ functions of the $R^2$ truncation}
In order to study the RG flow resulting from the ansatz \eqref{ansatz}, it is convenient to work with the dimensionless couplings
\be\label{dimless}
g_k = \tfrac{G_k}{k^{2-d}}  \, , \quad \lambda_k = \tfrac{\bar{\lambda}_k}{k^2} \, , \quad b_k = \tfrac{\bar{b}_k}{k^{4-d}} \, .
\ee
Their scale dependence is governed by the $\b$ functions
\be
\begin{split} \label{beta}
&  \p_t g_k = \beta_g(g, \lambda, b) \, , \\
&  \p_t \lambda_k = \beta_\lambda(g, \lambda, b) \, , \\
&  \p_t b_k = \beta_b(g, \lambda, b) ,
\end{split}
\ee
with $t \equiv \ln k$ being the RG time.
Their explicit form can be found in \cite{Lauscher:2002sq} to which we refer for further details. 

There the $\beta$ functions \eqref{beta} have been derived with the help of the transverse-traceless decomposition
of the fluctuation fields. This decomposition give rise to ``zero-mode'' contributions proportional to $\delta_{d,4}$.
A detailed investigation shows that these terms give rise to additional singularities of the $\b$ functions, which
also have a significant influence on the properties of the RG flow in the UV (see below for a more detailed discussion).
Since there is no reason to expect that similar singularities arise when the computation is carried out without the 
use of the transverse traceless decomposition, we remove these contributions in four dimensional calculations.
\footnote{Alternatively, this can be seen as analyzing the ``regularized'' system of $\b$ functions \eqref{beta} in $d=4-\epsilon$ dimensions.}


When studying the RG flow arising from \eqref{beta}, we first note 
\be\label{planes}
\beta_g(\lambda, g, b)|_{g = 0} = 0 \, , \qquad \beta_b(\lambda, g, b)|_{b = 0} = 0 \, , 
\ee
which implies that RG trajectories cannot cross the $g=0$ plane and the $b=0$ plane. As a consequence, a RG trajectory starting 
with a positive Newtons constant cannot pass to a regime with negative Newtons constant dynamically. Eq.\ \eqref{planes} does not ensure 
the positivity of $b_k$, however, since the coupling $b_k$ can change sign by passing $1/b_k = 0$, which is a regular point of the $\b$ functions. 

\begin{figure}[b!]
\begin{center}
\includegraphics[width=0.45\textwidth]{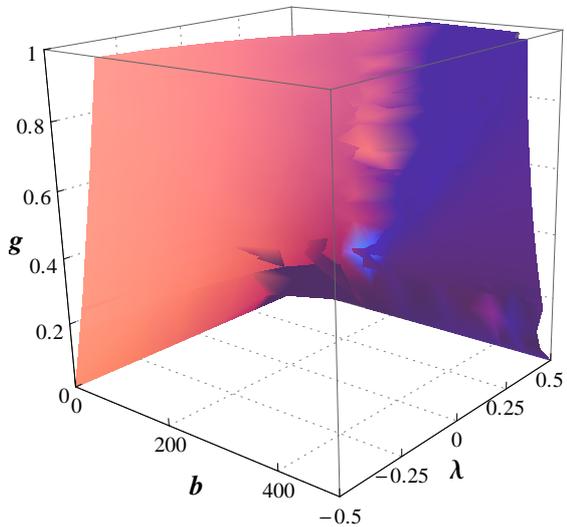}
\caption{The singular locus in theory space where the anomalous dimensions $\eta_N$ and $\eta_b$ diverge.}
\label{fig:poles}
\end{center}
\end{figure}

In addition, the $\b$ functions \eqref{beta} contain a singular locus where the denominators 
of the anomalous dimensions $\eta_N \equiv \p_t \ln g_k$ and $\eta_b \equiv \p_t \ln b_k$ vanishes. 
For $d=4$, the resulting singular locus is depicted in Fig.\ \ref{fig:poles}.

The final ingredient in understanding the RG flow of the theory are the fixed points where $\beta_\alpha(u^*) = 0$ for all $\alpha = 1, 2, \cdots$. In
the vicinity of such a fixed point of the RG equations \eqref{beta}, the linearized flow is governed by the Jacobi matrix
${\bf B}=(B_{\alpha \gamma})$, $B_{\alpha \gamma}\equiv\partial_\gamma  \beta_\alpha(u^*)$:
\begin{eqnarray}
\label{reu:H2}
k\,\partial_k\,{u}_\alpha(k)=\sum\limits_\gamma B_{\alpha \gamma}\,\left(u_\gamma(k)
-u_{\gamma}^*\right)\;.
\end{eqnarray}
The general solution to this equation reads
\be
\label{linflow}
u_\alpha(k)=u_{\alpha}^*+\sum\limits_I C_I\,V^I_\alpha\,
\left(\frac{k_0}{k}\right)^{\theta_I} \, ,
\ee
where the $V^I$'s are the right-eigenvectors of ${\bf B}$ with eigenvalues 
$-\theta_I$, i.e., $\sum_{\gamma} B_{\alpha \gamma}\,V^I_\gamma =-\theta_I\,V^I_\alpha$. Since ${\bf B}$ is not symmetric in general the $\theta_I$'s are not guaranteed to be real.

\subsection{The phase diagram in $d=4$}

In $d=4$, and for an exponential cutoff the fixed point structure of the system \eqref{beta} has already been analyzed in \cite{Lauscher:2002sq}. Besides the Gaussian fixed point (GFP) one encounters a 
NGFP  at 
$\{\lambda^*, g^*, b^*\} = \{ 0.330, 0.292, 183.5\}$ 
with three UV-relevant eigendirections
\be\label{FPexp}
\theta_{1,2} = 2.15 \pm 3.79 i \, , \qquad \theta_3 = 28.8 \, . 
\ee
In order to analyze the corresponding RG flow on phase space, we first repeat this analysis using the optimized cutoff. This also gives rise to a GFP and a NGFP.

\noindent
{\bf Gaussian Fixed Point} \\
Firstly, the system \eqref{beta} exhibits a GFP, 
\be\label{GFP4}
{\rm GFP}: \qquad \{\lambda^*, g^*, b^* \} = \{ 0 , 0 , 0 \}. 
\ee
Its stability coefficients are
\be
\{\theta_1, \theta_2, \theta_3\} = \{2, 0, -2\} \, ,
\ee
and thus correspond to the mass dimensions of the couplings. The linearized stability analysis shows that we have one IR-attractive, one IR-repulsive and one marginal eigendirection, in agreement with standard power counting. Since the GFP lies on the singular locus shown in Fig.\ \ref{fig:poles}, the corresponding stability matrix and its eigendirections depend on 
the precise order in which the limit $\{\lambda^*, g^*, b^* \} \rightarrow  \{ 0 , 0 , 0 \}$ is taken. To get a rough idea of the behavior close to the GFP one can build the following limits independent of the approaching direction
\begin{eqnarray}
\p_t \lambda_k|_{g = 0, b = 0} &=& -2 \lambda_k + \mathcal{O}(\lambda^2_k) \, ,\nonumber\\
\p_t g_k|_{\lambda = 0, b = 0} &=& 2 g_k + \mathcal{O}(g^2) \, ,\\
\p_t b_k|_{g = 0, \lambda = 0} &=& - \tfrac{1}{(4\pi)^2} \tfrac{419}{1080} b^2_k \, .\nonumber
\end{eqnarray}
Since $\p_t b_k$ is quadratic in $b$ it is IR-repulsive for positive $b$ and IR-attractive for negative $b$.

\noindent
{\bf Non-Gaussian Fixed Point} \\
In addition the four-dimensional RG flow gives rise to a unique NGFP, which is located at $\{ \lambda^* \, , \, g^* \, , \, b^* \,  \} = \{ 0.170,  0.754, 336.2 \}$. Its stability coefficients are
\be\label{NGFPd4}
\theta_1 = 3.95 \, , \qquad \theta_{2,3} =  1.56 \pm 3.31 i \, .
\ee
The positive real part of the critical exponents indicates that
any RG trajectory in the vicinity of the fixed point is
attracted to the NGFP in the UV while the non-vanishing 
imaginary part of $\theta_{2,3}$ shows that the trajectories will spiral into the NGFP. 

Notably, the critical exponent $\theta_1$ is significantly smaller than the one found in  eq.\ \eqref{FPexp}.
This difference can be traced back to the zero-mode contribution in the $\b$ functions. Including these 
the optimized cutoff actually gives rise to \emph{two} NGFPs
\be
\begin{split}
{\rm NGFP}_1: \; \{ \lambda^* \, , \, g^* \, , \, b^* \,  \} = \{ 0.163,  0.744, 345 \} \\
{\rm NGFP}_2: \; \{ \lambda^* \, , \, g^* \, , \, b^* \,  \} = \{ 0.173,  0.696, 455 \}
\end{split}
\ee
with
\be
\begin{split}
{\rm NGFP}_1: \; & \theta_1 = 3.49 \, , \quad  \theta_{2,3} = 1.65 \pm 3.10 i \, ,   \\ 
{\rm NGFP}_2: \; & \theta_1 = 25.6 \, , \quad  \theta_{2,3} = 1.40 \pm 2.78 \, . 
\end{split}
\ee
The critical exponents of the second one are very close to the one found in the analysis \cite{Lauscher:2002sq}. 
When removing the zero-mode contribution ${\rm NGFP}_2$ vanishes and the NGFP$_1$ falls on top of the NGFP \eqref{NGFPd4}. Remarkably it was found in \cite{Codello:2007bd} that the largest critical exponent at the NGFP is of order $25$ for the $R^2$ truncation and approximately $2$ when the $R^3$ term is taken into account as well. Thus our FP with the largest critical exponent being $\theta_1 = 3.95$ fits well to the higher order results.

\begin{figure}[t!]
\begin{center}
\includegraphics[width=0.45\textwidth]{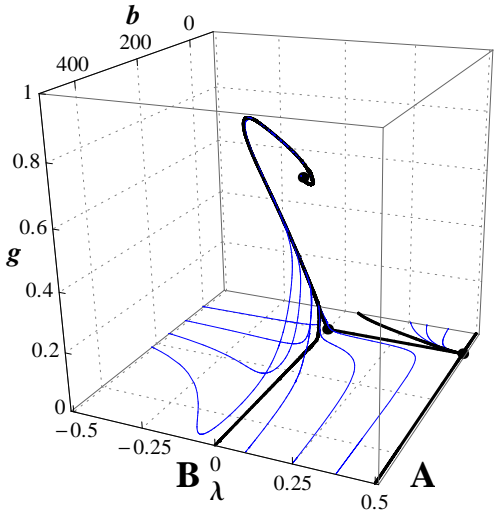}
\caption{A sample of possible trajectories including the examples A and B. The fixed points are marked with dots.}
\label{fig:flows4d}
\end{center}
\end{figure}

\noindent
{\bf Trajectories} \\
Starting at the NGFP there are several types of trajectories flowing toward the IR. In Fig.\ \ref{fig:flows4d}
we show a particular sample of them which exhibit a crossover to the GFP and are thus likely
to give rise to a classical regime.
In analogy to the phase diagram of the Einstein-Hilbert truncation depicted in Fig.\ \ref{fig:eh}, 
we distinguish RG trajectories of Type Ia and IIIa, according to their IR behavior as follows:
Upon their crossover, trajectories of Type Ia run toward negative values $\lambda$. For $k \rightarrow  0$, they 
approach the point $\{\lambda, g, b\}=\{-\infty, 0, 0\}$. They exhibit a classical regime with a negative value of the dimensionful cosmological
constant $\bar{\lambda}_k$. In contrast, the trajectories of Type IIIa leave the GFP regime in direction of positive cosmological constant. They may first flow
parallel to the $\lambda$ axis toward increasingly positive values $\lambda_k$. Here the trajectories
develop a classical regime with a positive cosmological constant $\bar{\lambda}_k$. Before
hitting the singular locus depicted in Fig.\ \ref{fig:poles}, however, the trajectories essentially turn
perpendicular to the $\lambda$-$g$-plane and flow toward increasing values $b_k$. This turn
can already occur at rather small values $\lambda_k$, such that the flow is along the $b$ axis.
Trading the coupling $b_k$ for $1/b_k$, 
shows that the $R^2$ coupling changes sign. At the end all these trajectories tend toward the point $\{\lambda, g, b\}=\{1/2, 0, 0^-\}$.
Note that these observations are in perfect agreement with the recent suggestions
\cite{Donkin:2012ud,Nagy:2012rn,Litim:2012vz} that the corresponding point $\{\lambda, g\}=\{1/2,0\}$
in the Einstein-Hilbert truncation actually constitutes an IR fixed point.

For later purposes, we marked two explicit Type IIIa sample trajectories {\bf A} and {\bf B}. 
These arise from the initial conditions
\be \label{example4d}
\begin{split}
{\rm \bf A}: \; & \left\{ \lambda_{\rm ini} \, , \, g_{\rm ini} \, , \, b_{\rm ini} \, \right\} = \left\{ 0.4999 \, , \, 10^{-10} \, , \, 10^9 \right\} \, , \\
{\rm \bf B}: \; & \left\{ \lambda_{\rm ini} \, , \, g_{\rm ini} \, , \, b_{\rm ini} \, \right\} = \left\{ 10^{-5} \, , \, 10^{-8} \, , \, 500 \right\} \, .
\end{split}
\ee
They will form the basis for our discussion of the spectral dimension in section \ref{sec:spectral}.

Finally, we remark that we have not been able to construct a RG trajectory that connects
the NGFP with the GFP and which would constitute the analogue of the separatrix (Type IIa trajectory) shown in Fig.\ \ref{fig:eh}.
The non-existence of a separatrix in the $R^2$ truncation is caused by the singularity structure of the $\b$ functions:
The RG flow away from the NGFP cannot be aligned to the IR-attractive eigendirection of the GFP. Thus we can finetune the flow such that the trajectory passes arbitrarily close to the GFP, 
but there is no solution connecting the fixed points.

\subsection{The phase diagram in $d=3$}

In this subsection we will repeat the discussion of the previous one for the case of three spacetime dimensions. The three dimensional case is a model very well suited for comparison to Monte Carlo data as e.g. \cite{Benedetti:2009ge} since the computations in three dimensions are less expensive within the CDT framework.

\noindent
{\bf Gaussian Fixed Point} \\
Again, the GFP is situated at $\{\lambda^*, g^*, b^* \} = \{ 0 , 0 , 0 \}$. Its associated stability coefficients are given by
\be
\{\theta_1, \theta_2, \theta_3\} = \{2, 1, -1\} \, .
\ee
These correspond to the mass-dimension of the (dimensionful) coupling constants. Eq.\ \eqref{linflow} then indicates that there are two IR-repulsive and one IR-attractive eigendirection. These depend on the direction of the approach to the GFP which is again situated on the singular locus. The limits
\begin{eqnarray}\label{eq:linear}
\p_t \lambda_k|_{g = 0, b = 0} &=& -2 \lambda_k + \mathcal{O}(\lambda^2_k) \, ,\nonumber\\
\p_t g_k|_{\lambda = 0, b = 0} &=& g_k + \mathcal{O}(g^2_k) \, ,\\
\p_t b_k|_{g = 0, \lambda = 0} &=& -b_k - \tfrac{1}{(4\pi)^2} \tfrac{79}{54} b_k^2 \, ,\nonumber
\end{eqnarray}
again are independent of the direction of the approach and give a rough idea of the behavior close to the GFP. This time the $b$ direction is IR-repulsive for positive and negative $b$ values.

\noindent
{\bf Non-Gaussian Fixed Points} \\
Investigating the limit $g = 0$, one first encounters a non-trivial FP at $\{\lambda^*, g^*, b^*\} = \{0, 0, - \tfrac{54}{79}(4\pi)^2\}$. This FP has two eigendirections which both have no $g$ component. Thus all trajectories connected to this FP can not leave the $g=0$ plane. We do not consider this FP any further.

In addition the system \eqref{beta} gives rise to a \emph{physical} NGFP which is located at $\{ \lambda^* , g^* , b^* \} = \{ 0.019 \, , \, 0.188 \, , \,  126.4 \, \}$. Its critical exponents read
\be\label{NGFPd3}
\{\theta_1, \theta_2, \theta_3\} = \{8.39, 1.86, 1.35\} \, .
\ee
Thus this NGFP comes with three UV attractive directions and all trajectories in its vicinity are dragged into this FP in the UV. The reality of the critical exponents indicates that the
 curling into the NGFP seen in four dimensions is absent. By varying the dimension in our theory one can see that the NGFPs \eqref{NGFPd3} and \eqref{NGFPd4} are continuously connected in $d$.

\begin{figure}[b!]
\begin{center}
\includegraphics[width=0.45\textwidth]{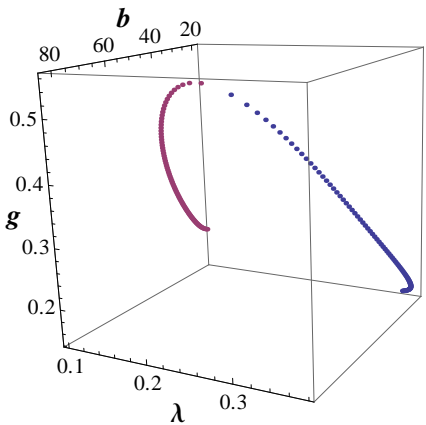}
\caption{Position of the UNGFP$_1$ (blue) and UNGFP$_2$ (red) depending on $d$. The two fixed points annihilate at $d=3.84$.}
\label{fig:3dto4d}
\end{center}
\end{figure}
Besides this NGFP one finds two additional \emph{unphysical} fixed points (UNGFPs). For completeness, their position and stability properties are collected in Table \ref{tab.fp1}.
\begin{table*}[t!]
\begin{center}
\begin{tabular}{|c|  c c c | c c c | }
\hline
  & \quad $\lambda^*$ \quad & \quad $g^*$ \quad & \quad $b^*$ \quad & \quad $\theta_1$ \quad & \quad $\theta_2$ \quad & \quad $\theta_3$ \quad  \\
\hline
UNGFP$_1$ &  0.364 &  0.147 & 18.167 & \multicolumn{2}{|c}{$1.56 \pm 4.84 i $} & - 9.85   \\
UNGFP$_2$ &  0.099 &  0.216 & 18.835 & \multicolumn{2}{c}{$0.19 \pm 0.97 i $} & 1.68     \\
\hline
\hline
\end{tabular}
\end{center}
\caption{\small Additional fixed points of the beta functions \eqref{beta} in $d=3$. These fixed points are labeled ``unphysical'' since the RG flow emanating from them does not give rise to a classical regime where General Relativity constitutes a good approximation.}
\label{tab.fp1}
\end{table*}
By tracing the flow of the fixed points UNGFP$_1$ and UNGFP$_2$ under an increase of the dimension $d$, one finds that they annihilate each other for $d=3.84$ at $\{\lambda^*, g^*, b^*\} = \{0.27, 0.54, 76.9\}$ as shown in Fig.\ \ref{fig:3dto4d}.  A similar continuous treatment of the dimension for the $\mathbb Z_2$-effective potential in \cite{Codello:2012sc} showed the appearance of new fixed points when decreasing the dimension $d$ as well.

Compared to the NGFP the UNGFPs are situated on the other side of the singularity locus. In principle, both UNGFPs do act as UV attractors for the RG trajectories spanning their UV-critical hypersurface. Tracing the RG flow toward the IR reveals, however, that the corresponding RG trajectories do not flow to a region in theory space, where General Relativity provides a good approximation. Thus, while still giving rise to RG trajectories which are asymptotically safe in the UV, they do not give rise to a classical regime. 

\begin{figure}[b!]
\begin{center}
\includegraphics[width=0.42\textwidth]{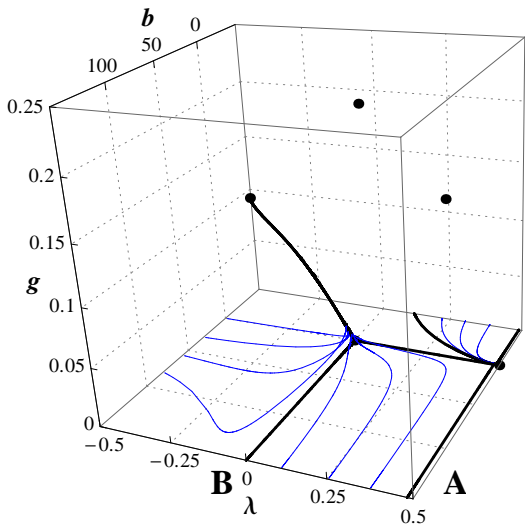}
\caption{Phase diagram in three dimensions with the GFP, the NGFP and the two UNGFPs. We depict a set of trajectories which develops a long classical regime and also includes the examples \eqref{example3d} highlighted with bold black lines.}
\label{fig:flows3d}
\end{center}
\end{figure}

\noindent
{\bf Trajectories} \\
In three dimensions the phase diagram is similar to the one in four dimensions. Again we focus on RG trajectories that exhibit a crossover from the NGFP to the GFP.
Some typical examples are depicted in Fig.\ \ref{fig:flows3d} and for later purposes we marked the trajectories \textbf A and \textbf B. These are obtained with the initial conditions
\be\label{example3d}
\begin{split}
{\rm \bf A}: \; & \left\{ \lambda_{\rm init} \, , \, g_{\rm init} \, , \, b_{\rm init} \, \right\} = \left\{ 0.49 \, , \, 10^{-8} \, , \, 5 \times 10^4 \right\} \, , \\
{\rm \bf B}: \; & \left\{ \lambda_{\rm init} \, , \, g_{\rm init} \, , \, b_{\rm init} \, \right\} = \left\{ 10^{-5} \, , \, 10^{-8} \, , \, 500 \right\} \, .
\end{split}
\ee
Besides the GFP and the NGFP, Fig.\ \ref{fig:flows3d} also depicts the two UNGFPs. These are separated from the NGFP by the singular locus shown 
in Fig.\  \ref{fig:poles}. Moreover, we highlighted the prospective IR fixed point $\{\lambda, g, b\} = \{ 0.5, 0, 0^- \} $.

The first obvious difference to the four dimensional plot in Fig.\ \ref{fig:flows4d} is that in three dimensions the approach toward the NGFP is straight since the critical exponents \eqref{NGFPd3} are real, compared to the complex ones \eqref{NGFPd4} in four dimensions. 

As we did in four dimensions we can divide the trajectories into Type Ia and Type IIIa trajectories. The former are going toward $\{\lambda, g, b\} = \{-\infty, 0, 0\}$ in the IR. The latter pass the GFP and head toward infinite $b$ values at positive $\lambda_k$ afterward. Again the use of $1/b$ instead of $b$ shows that the sign changes and the Type IIIa trajectories go toward the point $\{\lambda, g, b\}=\{1/2, 0, 0^- \}$ with negative $b$ values. 

In three (as well as four) dimensions we get large contributions to the $\b$ functions between the GFP and the NGFP as well as close to $\lambda = 1/2$. These are caused by the singularity structure of the truncated theory space and lead to a strong running in the $b$ direction.

\section{RG-improved diffusion processes}
\label{sec:diffusion}
As a first application, we use the phase-space study of the last section
to investigate the influence of higher-derivative terms on the
fractal-like properties of the effective QEG spacetimes. 
In this course, we start by adapting the ``RG improvement''
scheme for the diffusion of a test particle on the QEG spacetime \cite{Lauscher:2005qz,Reuter:2011ah}
to the higher-derivative action \eqref{ansatz}. 

The key idea of the RG improvement is to study diffusion processes with a spacetime
metric, which depends on the RG scale $k$
\be
\p_T K(x, x^\prime; T) = - \Delta(k) K(x, x^\prime; T) \, . 
\ee
Resorting to the ``flat-space'' approximation, the scale $k$ is identified with the momentum $p$ of
the plain waves used to probe the structures of spacetime and thus provides a (inverse) length scale at which the system is probed.
Taking the trace of $K(x, x^\prime; T)$, gives the average return probability
\be\label{retpro}
P(T) = \int \frac{d^dk}{(2\pi)^d} \e^{-k^2 F(k^2) T} \, , 
\ee
where the function $F(k^2)$ relates the Laplacian at a fixed reference scale $k_0$ (taken in the IR)
and the scale $k$: $\Delta(k) = F(k^2) \Delta(k_0)$. The scale-dependent spectral dimension is defined as the logarithmic $T$ derivative
\be\label{spectraldef}
D_s(T) \equiv -2 \frac{d \ln P_g(T)}{d \ln T} \, . 
\ee
For $F(k^2) = 1$ this relation reproduces the classical case $D_s(T) = d$; thus all quantum corrections
to the spectral dimension are encoded in $F(k^2)$.

Assuming that $F(k^2) \propto k^\delta$ undergoes power law scaling,
the integral \eqref{retpro} can be evaluated analytically, yielding
$P(T) \propto T^{-d/(2+\delta)}$. Substituting this relation into
\eqref{spectraldef} yields the spectral dimension
\be\label{fracdim}
\cD_s(T) = \frac{2d}{2+\delta} \, . 
\ee
For the Einstein-Hilbert case \cite{Reuter:2011ah}
\be\label{deh}
\delta^{\rm EH}(g, \lambda) = 2 + \lambda^{-1} \beta_\lambda(g, \lambda)
\ee
can be expressed as a function on theory space. The main task of this section is to derive
the analogue of this expression for the ansatz \eqref{ansatz}, taking the $R^2$ contribution
into account.

Following \cite{Reuter:2011ah}, we first write down the equations of motion arising from \eqref{ansatz}. Upon integrating by parts and dropping the surface terms these are
\begin{eqnarray} \label{eom}
&\left(- R + 2 \lb_k + \tfrac{16 \pi G_k R^2}{\bar{b}_k} \right) g^{\m\n}
+ \left( 2 - \tfrac{64 \pi G_k R}{\bar{b}_k} \right) R^{\m\n} \nonumber \\
& + \tfrac{64 \pi G_k}{\bar{b}_k} \left( D^\m D^\n R - (D^2 R) \, g^{\m\n}  \right) = 0 \, . 
\end{eqnarray}
In contrast to the classical equations of motion, the $k$ dependence of the coupling constants
promotes \eqref{eom} to a one-parameter family of equations of motion, each yielding
an effective description of the physics at the fixed scale $k$. In order to extract
the $k$ dependence of the metric, we first keep the scale $k$ fixed and substitute the ansatz 
\be \label{eq:eomSol}
R_{\m\n}\left( g|_k \right) = \frac{c_k}{d}  g_{\m\n}|_k \, ,
\ee
which implies constant curvature $R\left( g|_k \right)=c_k$. For this ansatz, the second line in \eqref{eom} is zero while the first line yields
\be\label{ceq}
2 \lb_k - \tfrac{1}{d} \left(d-2 \right) c_k + \tfrac{1}{d} \, \left(d-4\right) \, \tfrac{16 \pi G_k}{\bar{b}_k} \, c_k^2 = 0 \, .
\ee
At this stage, it is convenient to distinguish the two cases $d=4$ and $d \not = 4$.

For $d=4$ the term quadratic in $c_k$ vanishes and eq.\ \eqref{ceq} is easily solved
\be\label{ckd4}
c_k|_{d=4} = 4 \lb_k \, . 
\ee
Comparing solutions \eqref{eq:eomSol} at two different scales $k$ and $k_0$, where $k_0$ is a fixed reference scale taken in the classical regime, and using the identity $R^\m_\n(cg)=c^{-1}R^\m_\n(g)$, which holds for constant $c>0$, then yields
\be\label{confrela}
R_\m^\n\left( g_k \right) = R_\m^\n\left( \tfrac{c_k}{c_{k_0}} g_{k_0} \right) \, .
\ee
This equation allows to read off the function $F(k^2)$ relating the metrics at the scale $k$ and $k_0$
\be\label{confrel}
g^{\m\n}|_k = F(k) \, g^{\m\n} |_{k_0} \, ,
\ee
with $F(k) = \bar{\lambda}_k/\bar{\lambda}_{k_0}$. In a regime where $F(k^2$ undergoes power-law scaling for an extended $k$-interval, the function $\delta(k)$ can then be obtained as $\delta(k) = k \p_k \ln F(k)$ and reads
\begin{equation} \label{eq:delta4d}
\delta(g, \lambda, b)|_{d=4} = 2 + \lambda^{-1} \beta_\lambda \, .
\end{equation}
Formally, this result has the same form as in the Einstein-Hilbert case \eqref{deh}. For the $R^2$ truncation the $\b$ function $\beta_\lambda$ depends on the three couplings $\lambda, g, b$, however, so that there are non-trivial corrections.

For $d \not = 4$ the situation is slightly more complicated. In this case the
 solutions of the quadratic equation are 
\be\label{csol}
c_k^{\pm} = \tfrac{(d-2) \bar b_k}{32 \pi (d-4) G_k} \, \left( \, 1 \pm \sqrt{ 1 - \tfrac{h_d \, G_k \, \lb_k}{\bar b_k} }  \, \right) \, ,
\ee
where
\be
h_d = \frac{128 d (d-4) \pi}{(d-2)^2} \, .
\ee

Investigating the conditions implied by eq.\ \eqref{confrela}, one then observes that $c_k^\pm$ may not be positive definite. In particular for $d=3$ and positive $G$ and $\bar b$, $c_k^+$ is negative. Therefore we discard this solution and focus on $c_k^-$. Notably, $c_k^-$ also has a well-defined limit $d \rightarrow 4$ where it reduces to \eqref{ckd4}. Thus selecting the $c_k^-$ branch will lead to a spectral dimension which is continuous in $d$.

The function $F(k^2)$ originating from $c_k^-$ can then be obtained completely analogous to the case $d=4$ and reads
\be\label{Fgen}
F(k) = \frac{G_0 \bar{b}_k}{G_k \bar{b}_0}  \, \frac{1 - \sqrt{1- h_d G_k \bar{\lambda}_k / \bar{b}_k}}{1 - \sqrt{1- h_d G_0 \bar{\lambda}_0 / \bar{b}_0} } \, .
\ee
Again, this leads to a  $\delta(k)$ which is given by a function on theory space
\be\label{deltamod}
\delta = 2 - \frac{\beta_g}{g} + \frac{\beta_b}{b} + \frac{\tfrac{h_d}{2}\left( \tfrac{g}{b} \beta_\lambda - \tfrac{g \lambda}{b^2} \beta_b + \tfrac{\lambda}{b} \beta_g \right)}{\left( 1 - \sqrt{1- h_d g \lambda / b} \right)\sqrt{1- h_d g \lambda / b }} \, .
\ee
Substituted into the spectral dimension \eqref{fracdim}, this result captured
the $R^2$ corrections to $D_s$. Its properties will be investigated in the next section.
Keep in mind, however, that $D_s$ is ill-defined if $\delta$ changes rapidly \cite{Reuter:2011ah}.

\section{Fractal properties of QEG spacetimes}
\label{sec:spectral}
In this section, we analyze the spectral dimension obtained along
the sample trajectories \eqref{example4d} and \eqref{example3d} in $d=4$ and $d=3$, respectively.
For this purpose we evaluate \eqref{deltamod} along the trajectory and substitute
the result into \eqref{fracdim} to obtain the spectral dimension $D_s(t)$ depending 
on the logarithmic RG time $t = \ln k$.

\noindent
{\bf Spectral Dimension in $d=4$} \\
For $d=4$ the two exemplary trajectories \textbf A and \textbf B are depicted in Fig.\ \ref{fig:flows4d} .
Both trajectories emanate from the NGFP in the UV and flow toward the GFP.
After passing the GFP trajectory \textbf A flows toward $\lambda = 1/2$ along the $\lambda$ axis
before running toward large values $b_k$. Trajectory \textbf B leaves the GFP regime close to the $b$ axis instead. 

\begin{figure}[b!]
\begin{center}
\includegraphics[width=0.4\textwidth]{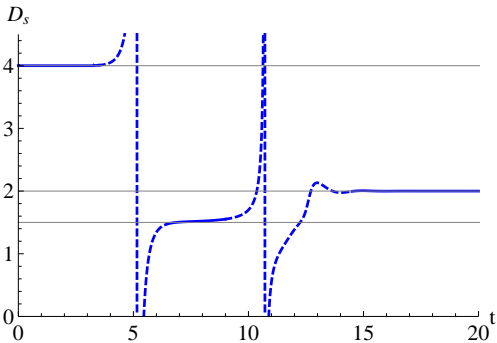}
\caption{Spectral dimension $D_s$ depending on RG time $t$ for the trajectory \textbf A in $d=4$.}
\label{fig:DsA4d}
\end{center}
\end{figure}

The spectral dimension along trajectory \textbf A is shown in Fig.\ \ref{fig:DsA4d}.
The diagram displays three distinguished plateaus marked by the solid blue lines.
From the UV (large $t$) to the IR (small $t$) these are the NGFP plateau with
$D_s = 2$, the semi-classical plateau with $D_s \simeq 1.5$ and the classical plateau
where $D_s =4$, respectively. They are connected by interpolating regimes (dashed), where $\delta(t)$ changes
rapidly and thus $D_s(t)$ is not well defined. The poles in these regions are caused by
$\delta(t) \rightarrow -2$, which in turn is linked to $\lambda_k$ changing sign when 
passing from the NGFP to the GFP.

Surprisingly, the spectral dimension of the semi-classical plateau, $D_s \simeq 1.5$
does not agree with the one observed in the Einstein-Hilbert truncation $D_s^{\rm EH} = 4/3$ \cite{Reuter:2011ah}.
This difference can be traced back to the fact, that in the $R^2$ truncation the 
 flow for this intermediate part of the trajectory is dominated by the singular 
 locus Fig.\ \ref{fig:poles}. This locus induces a scaling behavior, $k^\delta$, $\delta \simeq 3.3$, different from the $k^4$-scaling
 associated to the GFP.

\begin{figure}[b!]
\begin{center}
\includegraphics[width=0.4\textwidth]{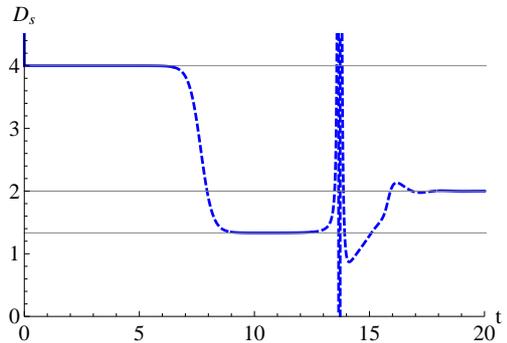}
\caption{Spectral dimension $D_s$ depending on RG time $t$ for the trajectory \textbf B in $d=4$.}
\label{fig:DsB4d}
\end{center}
\end{figure}

The spectral dimension of trajectory {\bf B}, shown in Fig.\ \ref{fig:DsB4d} also exhibits
three plateaus which are connected by short interpolating pieces where $\delta(t)$ changes 
rapidly.
 The NGFP plateau ($D_s=2$) and the classical plateau ($D_s=4$) have the same spectral dimension as for the trajectory {\bf A}.
In this case, the semi-classical plateau forms at $D_s=4/3$, however, and thus has the same value as in the Einstein-Hilbert truncation.
In terms of the underlying RG flow it is created by the trajectory flowing away from the GFP along the $b$ axis (see the Appendix for an analytic investigation). 
Notably, there is no plateau corresponding to the scaling caused by the singular locus. This feature is shadowed by the poles where $\delta \rightarrow -2$.

\noindent
{\bf Spectral Dimension in $d=3$} \\
When investigating the spectral dimension in $d=3$, we first notice that the functional form of
\eqref{deltamod} is quite different from the four-dimensional case. In particular $\delta(\lambda, g, b)$ now depends 
on the $\b$ functions of all three couplings. In order to see if the new structures still support
a multi-fractal picture of spacetime, we again analyze $D_s(t)$ along the two exemplary trajectories \textbf A and \textbf B given in Fig.\ \ref{fig:flows3d}.

\begin{figure}[b!]
\begin{center}
\includegraphics[width=0.4\textwidth]{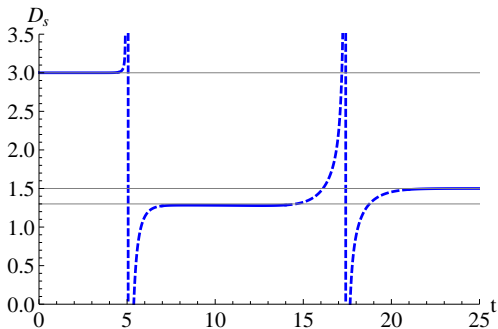}
\caption{Spectral dimension $D_s$ depending on RG time $t$ for the trajectory \textbf A in $d=3$.}
\label{fig:DsA3d}
\end{center}
\end{figure}

The spectral dimension resulting from trajectory \textbf A is depicted in Fig.\ \ref{fig:DsA3d}.
Also here we observe three plateaus (solid line) which are connected by short interpolating pieces (dashed line)
where the spectral dimension is not well-defined. In the IR and in the UV $D_s(t)$ again has the expected behavior:
at small RG times $t$ it equals the spacetime dimension three while for large RG times the NGFP induces 
a spectral dimension which is half of the spacetime dimension, $D_s=3/2$.

The semi-classical plateau has $D_s \simeq 1.3$ and again develops during the crossover between the NGFP and the GFP.
The poles framing the plateau are caused by $\lambda_k$ changing sign which again leads to $\delta = -2$. Similarly to the four-dimensional case,
the value of the semi-classical plateau does not correspond to the one observed in the Einstein-Hilbert case $D_s^{\rm EH} = 1$. Instead of a property of the GFP this plateau 
is created by scaling induced by the singular locus and thus has a quite different origin.

\begin{figure}[b!]
\begin{center}
\includegraphics[width=0.4\textwidth]{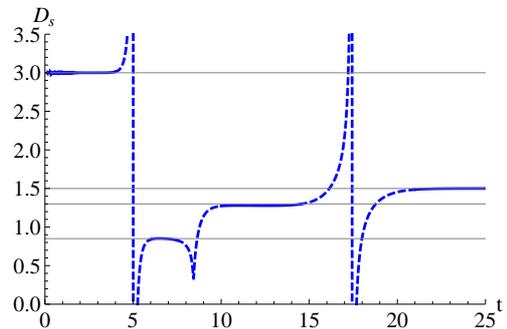}
\caption{Spectral dimension $D_s$ depending on RG time $t$ for the trajectory \textbf B in $d=3$.}
\label{fig:DsB3d}
\end{center}
\end{figure}

A special feature occurs when studying the spectral dimension along trajectory {\bf B}, which is shown in Fig.\ \ref{fig:DsB3d}. 
Besides the classical and NGFP plateaus, this trajectory develops {\it two} semi-classical regimes:
The plateau with $D_s \simeq 1.3$ is caused by the trajectory passing close to the singular locus when flowing
from the NGFP to the GFP. The second plateau with $D_s \simeq 0.85$ appears during the flow away from the GFP close to the $b$ axis
toward the classical regime. The latter plateau is the one that, in $d=4$, corresponds to the semi-classical plateau found in
the Einstein-Hilbert truncation. Thus it is natural to identify the shift $D_s^{\rm EH} = 1$ $\rightarrow$ $D_s^{R^2} \simeq 0.85$
as the correction induced by the inclusion of the $R^2$-terms in the computation.

\begin{figure}[t!]
\begin{center}
\includegraphics[width=0.4\textwidth]{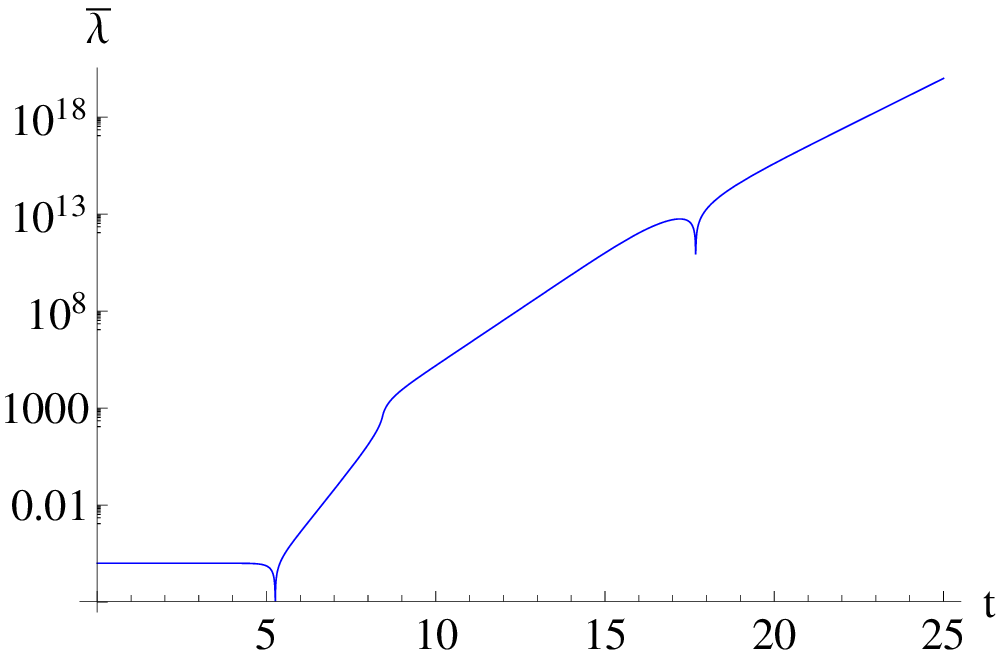} \hspace{2cm}
\includegraphics[width=0.4\textwidth]{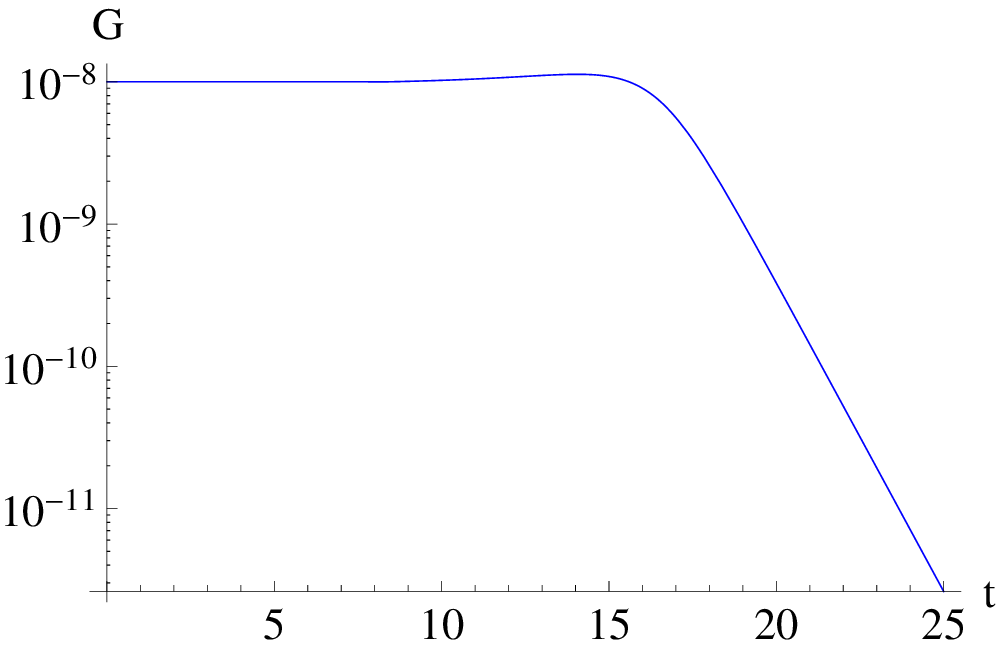}
\caption{Absolute value of dimensionful cosmological constant $\bar\lambda$ and dimensionful Newton constant $G$ depending on logarithmic RG scale $t$ along the exemplary trajectory {\bf B}.}
\label{fig:dimfullRunning}
\end{center}
\end{figure}

The four different scaling regimes underlying Fig.\ \ref {fig:DsB3d} are also visible in the running of the absolute value of the  dimensionful cosmological constant $\bar\lambda$ in Fig.\ \ref{fig:dimfullRunning} whereas the running of the dimensionful Newton constant shows only two distinct scaling regimes. Moreover the running of $\bar\lambda$ shows two peaks indicating the change of the sign. 
%

\section{Conclusion}
\label{sec:conclusion}
In this work, we used the $\b$ functions \cite{Lauscher:2002sq} to construct
the phase diagram of Quantum Einstein Gravity (QEG) in the $R^2$ truncation.
For $d=4$ and $d=3$ the resulting RG trajectories are shown in Figs.\ \ref{fig:flows4d} and \ref{fig:flows3d}, respectively.
Notably, the flow is governed by the interplay of its fixed points. In this sense the phase diagram of the $R^2$ truncation
is quite similar to the one of the Einstein-Hilbert truncation: 
the trajectories with a long classical regime result from a crossover from the 
non-Gaussian fixed point (NGFP) controlling the UV behavior of the theory to the Gaussian fixed point (GFP)
responsible for its classical properties.  Remarkably, the value 
of the $R^2$ coupling is easily compatible with experimental bounds \cite{Berry:2011pb,DeFelice:2010aj}. 

As an important new feature, our phase diagram is molded not only by its fixed points, but also by singular loci where
the anomalous dimensions diverge. The two regimes where this new mechanism is at work are the crossover between the NGFP and GFP
and the flow close to $\lambda = 1/2$. In both cases the singularities induce a strong running of the $R^2$-coupling $b$. Especially
close to the $\lambda =1/2$ locus the RG flow is essential parallel to the $\lambda =1/2$ boundary and the strong $b$ running allows to avoid 
this singularity. The resulting flow pattern closely resembles the one expected from the IR fixed point recently discussed
in \cite{Donkin:2012ud,Nagy:2012rn,Litim:2012vz}.  

As a first application of our classification, we studied the spectral dimension $D_s$ of the effective QEG spacetimes.
This is the first instance where higher-derivative corrections are taken into account
in an RG-improvement scheme. 
As a central result, we establish that the multi-fractal structure of the effective spacetimes \cite{Reuter:2011ah}, reflected by several
plateau values for $D_s$, is robust upon including $R^2$ effects. All our examples give $D_s = d$ in the classical regime and $D_s =d/2$ in the UV. The latter constitutes
the universal signature of the NGFP in the spectral dimension \cite{Lauscher:2005qz}.

Between the classical and UV regime we obtain the ``semi-classical'' plateau which, in the Einstein-Hilbert case, is closely linked to
universal properties of the RG flow close to the GFP and situated at $D_s = 4/3$ $(d=4)$ and $D_s = 1$ $(d=3)$. The $R^2$ construction
reveals a second mechanism that can lead to a similar ``semi-classical'' plateau: scaling induced by singularities. For our examples
this induces semi-classical plateaus with $D_s \approx 3/2$ $(d=4)$ and $D_s \approx 1.3$ $(d=3)$. Thus the semi-classical regime is sensitive to 
the details of the underlying RG trajectory. In this light, it is not surprising that different Monte-Carlo studies of the gravitational
path integral report different plateau values for $D_s$ at intermediate scales \cite{Benedetti:2009ge,Coumbe:2012qr}.

\section*{Acknowledgments}
We thank M.\ Reuter for many helpful and inspiring discussions and A.\ Eichhorn and G.\ Calcagni for comments on the manuscript. This work is supported by the Deutsche Forschungsgemeinschaft (DFG) within the Emmy-Noether program (Grant SA/1975 1-1).
\begin{appendix}
\begin{section}{The GFP in $d=4$}
In this section we investigate the GFP in four spacetime dimensions a little bit further. Since we are interested in the behavior of the trajectory B in Fig.\ \ref{fig:poles} we evaluate the stability matrix by first taking the limits $g \rightarrow 0$ and $\lambda \rightarrow 0$ and at the end the limit $b \rightarrow 0$. This procedure leads to the stability matrix
\be
B = \begin{pmatrix}
-2 & \tfrac{1}{2\pi} & 0 \\
0 & 2 & 0 \\
0 & 0 & 0
\end{pmatrix} \, .
\ee
One of the corresponding eigenvectors is the unit vector in $b$ direction and thus we are able to investigate the behavior along this direction. The upper left block of this stability matrix is exactly the stability matrix of the Einstein--Hilbert truncation and thus the solutions to the linearized flow equations are known to be
\begin{eqnarray}
\lambda_k &=& \left( \lambda_{k_0} - \tfrac{4}{2\pi} g_{k_0} \right) \left( \tfrac{k_0}{k} \right)^2 + \tfrac{4}{2\pi} g_{k_0} \left( \tfrac{k}{k_0} \right)^2 \, ,\nonumber \\
g_k &=& g_{k_0} \left( \tfrac{k}{k_0} \right)^2 \, .
\end{eqnarray}
Thus for the dimensionful cosmological constant $\bar\lambda$ we find
\be
\bar\lambda_k = \bar\lambda_{k_0} + \tfrac{4 G_{k_0}}{2\pi}\left( k^4 - k_0^4 \right) \, .
\ee
Since $\bar\lambda$ shows a $k^4$ scaling $\delta$ of \eqref{eq:delta4d} is four as it was already derived in \cite{Reuter:2011ah}. Thus we get the spectral dimension
\be
D_s = \tfrac{2d}{2 + \delta} = \tfrac{4}{3} \, .
\ee
\end{section}
\end{appendix}


\begin{thebibliography}{99}

\bibitem{Weinberg:1980gg} 
  S.~Weinberg
in \textit{General Relativity, an Einstein Centenary Survey},
S.W.~Hawking and W.~Israel (Eds.),
Cambridge University Press, 1979; \\
S.~Weinberg,
hep-th/9702027.

\bibitem{Weinproc1}
S.~Weinberg, arXiv:0903.0568; PoS C {D09} (2009) 001, arXiv:0908.1964.

\bibitem{Niedermaier:2006wt} 
  M.~Niedermaier and M.~Reuter,
  Living Rev.\ Rel.\  {\bf 9}, 5 (2006).

\bibitem{robrev}
R.~Percacci, in \textit{Approaches to Quantum Gravity: Towards a New Understanding of Space, Time and Matter}, D. Oriti (Ed.), Cambridge University Press, Cambridge, 2009,
  arXiv:0709.3851.


\bibitem{Reuter:2012id}
  M.~Reuter and F.~Saueressig,
   New J. Phys. 14 (2012) 055022,
  arXiv:1202.2274.

\bibitem{Reuter:1996cp} 
  M.~Reuter,
  Phys.\ Rev.\ D {\bf 57}, 971 (1998),
  hep-th/9605030.


\bibitem{souma}
W.~Souma,
Prog.\ Theor.\ Phys.\ 102 (1999) 181, hep-th/9907027.

\bibitem{oliver1}
O.~Lauscher and M.~Reuter,
Phys.\ Rev.\ D 65 (2002) 025013, hep-th/0108040.
%
\bibitem{frank1}
M.~Reuter and F.~Saueressig,
Phys.\ Rev.\ D 65 (2002) 065016, hep-th/0110054.


\bibitem{oliver3}
O.~Lauscher and M.~Reuter,
Class.\ Quant.\ Grav.\ 19 (2002) 483, hep-th/0110021.

\bibitem{Lauscher:2002sq} 
  O.~Lauscher and M.~Reuter,
  Phys.\ Rev.\ D {\bf 66}, 025026 (2002),
  hep-th/0205062.


\bibitem{Codello:2007bd} 
  A.~Codello, R.~Percacci and C.~Rahmede,
  Int.\ J.\ Mod.\ Phys.\ A {\bf 23}, 143 (2008),
  arXiv:0705.1769;
  P.~F.~Machado and F.~Saueressig,
  Phys.\ Rev.\ D {\bf 77}, 124045 (2008),
  arXiv:0712.0445.

\bibitem{Benedetti:2009rx} 
  D.~Benedetti, P.~F.~Machado and F.~Saueressig,
  Mod.\ Phys.\ Lett.\ A {\bf 24}, 2233 (2009),
  arXiv:0901.2984;


\bibitem{Fischer:2006at} 
  P.~Fischer and D.~F.~Litim,
  AIP Conf.\ Proc.\  {\bf 861}, 336 (2006),
  hep-th/0606135;
  Phys.\ Lett.\ B {\bf 638}, 497 (2006),
  hep-th/0602203.


\bibitem{Eichhorn:2009ah} 
  A.~Eichhorn, H.~Gies and M.~M.~Scherer,
  Phys.\ Rev.\ D {\bf 80}, 104003 (2009),
  arXiv:0907.1828;
  K.~Groh and F.~Saueressig,
  J.\ Phys.\ A A {\bf 43}, 365403 (2010),
  arXiv:1001.5032;
  A.~Eichhorn and H.~Gies,
  Phys.\ Rev.\ D {\bf 81}, 104010 (2010),
  arXiv:1001.5033.
  
\bibitem{Becker:2012js} 
  D.~Becker and M.~Reuter,
  arXiv:1205.3583.

\bibitem{Manrique:2011jc} 
  E.~Manrique, S.~Rechenberger and F.~Saueressig,
  Phys.\ Rev.\ Lett.\  {\bf 106}, 251302 (2011),
  arXiv:1102.5012.
  
\bibitem{Benedetti:2010nr} 
  D.~Benedetti, K.~Groh, P.~F.~Machado and F.~Saueressig,
  JHEP {\bf 1106}, 079 (2011),
  arXiv:1012.3081.


\bibitem{Litim:2012vz}
  D.~Litim and A.~Satz,
  arXiv:1205.4218.
  
\bibitem{Reuter:2001ag} 
  M.~Reuter and F.~Saueressig,
  Phys.\ Rev.\ D {\bf 65}, 065016 (2002),
  hep-th/0110054.
  
\bibitem{h3}
M.~Reuter and H.~Weyer,
JCAP\ 12 (2004) 001,
hep-th/0410119.  
  
   
\bibitem{Donkin:2012ud} 
  I.~Donkin and J.~M.~Pawlowski,
  arXiv:1203.4207.

\bibitem{Nagy:2012rn} 
  S.~Nagy, J.~Krizsan and K.~Sailer,
  arXiv:1203.6564.


\bibitem{reu:bh}
A.~Bonanno and M.~Reuter,
Phys.\ Rev.\ D 62 (2000) 043008, hep-th/0002196;
Phys.\ Rev.\ D 73 (2006) 083005, hep-th/0602159;
Phys.\ Rev.\ D 60 (1999) 084011, gr-qc/9811026.
%

\bibitem{reu:erick1}
M.~Reuter and E.~Tuiran, Phys.\ Rev.\ D 83 (2011) 044041, arXiv:1009.3528.


\bibitem{reu:cosmo1}
A.~Bonanno and M.~Reuter,
Phys.\ Rev.\ D 65 (2002) 043508, hep-th/0106133;\\ M.~Reuter and 
F.~Saueressig, JCAP\ 09 (2005) 012, hep-th/0507167.

\bibitem{reu:entropy}
A.\ Bonanno and M.\ Reuter, Journal of Phys.\ Conf.\ Ser.\ 140 (2008) 012008, arXiv:0803.2546;
JCAP {08} (2007) 024, arXiv:0706.0174; Entropy 13 (2011) 274, arXiv:1011.2794.

%
\bibitem{reu:cosmo2}
A.~Bonanno and M.~Reuter,
Phys.\ Lett.\ B 527 (2002) 9, astro-ph/0106468; \\ 
Int.\ J.\ Mod.\ Phys.\ D 13 (2004) 107, astro-ph/0210472.
%
\bibitem{reu:elo}
E.~Bentivegna, A.~Bonanno and M.~Reuter,
JCAP 01 (2004) 001, 
astro-ph/0303150.
%

%
\bibitem{reu:wein3}
S.~Weinberg, Phys.\ Rev.\ D 81 (2010) 083535, arXiv:0911.3165.
%
\bibitem{reu:h1}
M.~Reuter and H.~Weyer,
Phys.\ Rev.\ D 69 (2004) 104022,
 hep-th/0311196.


\bibitem{reu:Ward:2008sm}
  B.~F.~L.~Ward,
  Mod.\ Phys.\ Lett.\ A {23} (2008) 3299,
  arXiv:0808.3124.

\bibitem{reu:Bonanno:2010bt}
  A.~Bonanno, A.~Contillo and R.~Percacci,
  Class.\ Quant.\ Grav.\  {28} (2011) 145026,
  arXiv:1006.0192.

\bibitem{Hindmarsh:2011hx}
  M.~Hindmarsh, D.~Litim and C.~Rahmede,
  JCAP {07} (2011) 019,
  arXiv:1101.5401.
%
\bibitem{reu:h2}
M.~Reuter and H.~Weyer,
Phys.\ Rev.\ D 70 (2004) 124028, hep-th/0410117.
%
  
   
\bibitem{Lauscher:2005qz}
  O.~Lauscher and M.~Reuter,
  JHEP {\bf 0510} (2005) 050,
  hep-th/0508202.

\bibitem{Reuter:2011ah}
  M.~Reuter and F.~Saueressig,
  JHEP {12} (2011) 012, arXiv:1110.5224.

\bibitem{Ambjorn:2005db} 
  J.~Ambjorn, J.~Jurkiewicz and R.~Loll,
  Phys.\ Rev.\ Lett.\  {\bf 95}, 171301 (2005),
  hep-th/0505113;
  Phys.\ Rev.\ D {\bf 72}, 064014 (2005),
  hep-th/0505154.

\bibitem{Coumbe:2012qr} 
    J.~Laiho and D.~Coumbe,
  Phys.\ Rev.\ Lett.\  {\bf 107} (2011) 161301,
  arXiv:1104.5505.
  
  
\bibitem{Modesto:2008jz} 
  L.~Modesto,
  Class.\ Quant.\ Grav.\  {\bf 26}, 242002 (2009),
  arXiv:0812.2214;
  F.~Caravelli and L.~Modesto,
  arXiv:0905.2170;
  E.~Magliaro, C.~Perini and L.~Modesto,
  arXiv:0911.0437.

\bibitem{Modesto:2009qc}
  L.~Modesto and P.~Nicolini,
  Phys.\ Rev.\ D {\bf 81} (2010) 104040,
  arXiv:0912.0220.


\bibitem{Giasemidis:2012qk} 
  G.~Giasemidis, J.~F.~Wheater and S.~Zohren,
  arXiv:1202.2710;
  arXiv:1202.6322.
  
\bibitem{Calcagni:2011kn} 
  G.~Calcagni,
  arXiv:1106.5787;
  JHEP {\bf 1201}, 065 (2012),
  arXiv:1107.5041.

\bibitem{Calcagni:2009kc} 
  G.~Calcagni,
  Phys.\ Rev.\ Lett.\  {\bf 104}, 251301 (2010),
  arXiv:0912.3142;
  JHEP {\bf 1003}, 120 (2010),
  arXiv:1001.0571;
  Phys.\ Lett.\ B {\bf 697}, 251 (2011),
  arXiv:1012.1244.

\bibitem{Arzano:2011yt} 
  M.~Arzano, G.~Calcagni, D.~Oriti and M.~Scalisi,
  Phys.\ Rev.\ D {\bf 84}, 125002 (2011),
  arXiv:1107.5308.

\bibitem{Litim:opt} 
  D.~F.~Litim,
  Phys.\ Rev.\ D {\bf 64}, 105007 (2001),
  hep-th/0103195;
  Nucl.\ Phys.\ B {\bf 631}, 128 (2002),
  hep-th/0203006.

\bibitem{Benedetti:2009ge} 
  D.~Benedetti and J.~Henson,
  Phys.\ Rev.\ D {\bf 80}, 124036 (2009),
  arXiv:0911.0401.
  
\bibitem{Codello:2012sc} 
  A.~Codello,
  arXiv:1204.3877.

\bibitem{Berry:2011pb}
  C.~P.~L.~Berry and J.~R.~Gair,
  Phys.\ Rev.\ D {\bf 83} (2011) 104022,
  arXiv:1104.0819.

\bibitem{DeFelice:2010aj}
  A.~De Felice and S.~Tsujikawa,
  Living Rev.\ Rel.\  {\bf 13} (2010) 3,
  arXiv:1002.4928.

\end{thebibliography}
\end{document}